\documentclass[twoside,twocolumn,9pt]{article}
\usepackage{extsizes}
\usepackage[super,sort&compress,comma]{natbib} 
\usepackage[version=3]{mhchem}
\usepackage[left=1.5cm, right=1.5cm, top=1.785cm, bottom=2.0cm]{geometry}
\usepackage{balance}
\usepackage{times,mathptmx}
\usepackage{sectsty}
\usepackage{graphicx} 
\usepackage{lastpage}
\usepackage[format=plain,justification=justified,singlelinecheck=false,font={stretch=1.125,small,sf},labelfont=bf,labelsep=space]{caption}
\usepackage{float}
\usepackage{fancyhdr}
\usepackage{fnpos}
\usepackage[english]{babel}
\usepackage{booktabs}
\usepackage{textcomp}
\usepackage{gensymb}
\addto{\captionsenglish}{%
  
}
\usepackage{array}
\usepackage{droidsans}
\usepackage{charter}
\usepackage[T1]{fontenc}
\usepackage[usenames,dvipsnames]{xcolor}
\usepackage{setspace}
\usepackage[compact]{titlesec}
\usepackage{hyperref}
\usepackage{epstopdf}%This line makes .eps figures into .pdf - please comment out if not required.

\definecolor{cream}{RGB}{222,217,201}

\begin{document}

\pagestyle{fancy}
\thispagestyle{plain}
\fancypagestyle{plain}{

%%%HEADER%%%
\fancyhead[C]{\includegraphics[width=18.5cm]{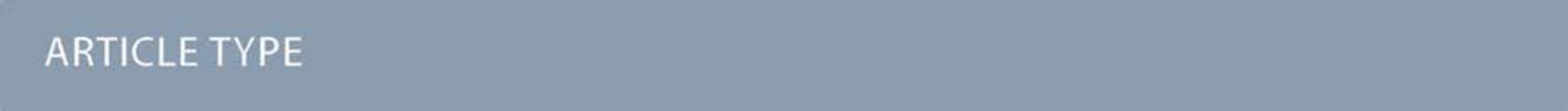}}
\fancyhead[L]{\hspace{0cm}\vspace{1.5cm}\includegraphics[height=30pt]{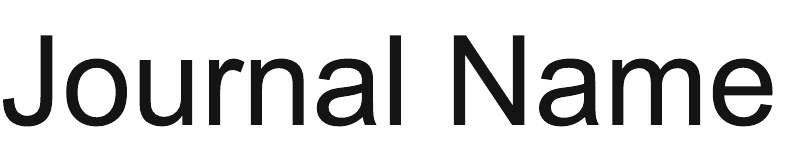}}
\fancyhead[R]{\hspace{0cm}\vspace{1.7cm}\includegraphics[height=55pt]{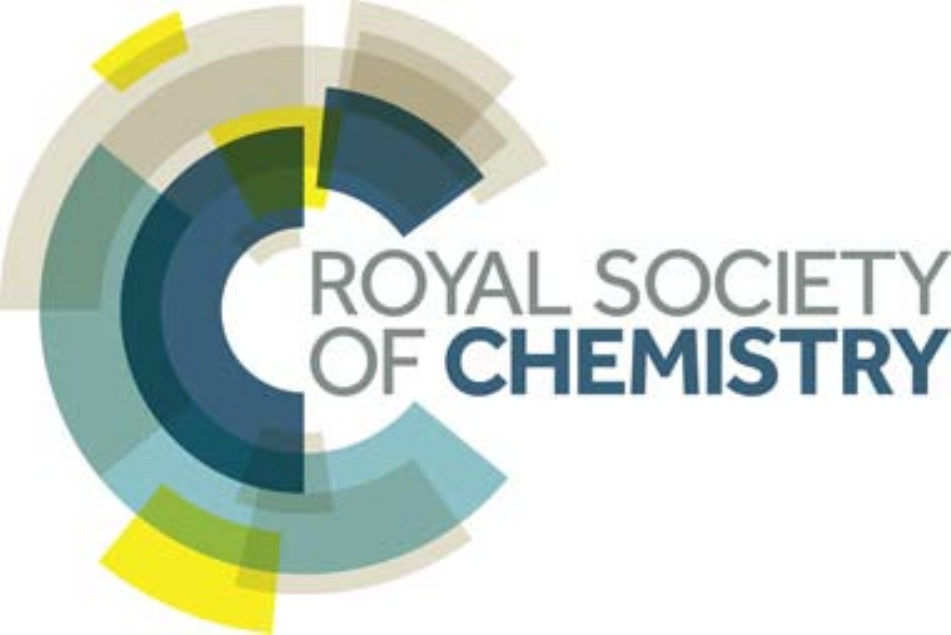}}
\renewcommand{\headrulewidth}{0pt}
}
%%%END OF HEADER%%%

%%%PAGE SETUP - Please do not change any commands within this section%%%
\makeFNbottom
\makeatletter
\renewcommand\LARGE{\@setfontsize\LARGE{15pt}{17}}
\renewcommand\Large{\@setfontsize\Large{12pt}{14}}
\renewcommand\large{\@setfontsize\large{10pt}{12}}
\renewcommand\footnotesize{\@setfontsize\footnotesize{7pt}{10}}
\makeatother

\renewcommand{\thefootnote}{\fnsymbol{footnote}}
\renewcommand\footnoterule{\vspace*{1pt}% 
\color{cream}\hrule width 3.5in height 0.4pt \color{black}\vspace*{5pt}} 
\setcounter{secnumdepth}{5}

\makeatletter 
\renewcommand\@biblabel[1]{#1}            
\renewcommand\@makefntext[1]% 
{\noindent\makebox[0pt][r]{\@thefnmark\,}#1}
\makeatother 
\renewcommand{\figurename}{\small{Fig.}~}
\sectionfont{\sffamily\Large}
\subsectionfont{\normalsize}
\subsubsectionfont{\bf}
\setstretch{1.125} %In particular, please do not alter this line.
\setlength{\skip\footins}{0.8cm}
\setlength{\footnotesep}{0.25cm}
\setlength{\jot}{10pt}
\titlespacing*{\section}{0pt}{4pt}{4pt}
\titlespacing*{\subsection}{0pt}{15pt}{1pt}
%%%END OF PAGE SETUP%%%

%%%FOOTER%%%
\fancyfoot{}
\fancyfoot[LO,RE]{\vspace{-7.1pt}\includegraphics[height=9pt]{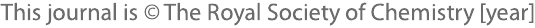}}
\fancyfoot[CO]{\vspace{-7.1pt}\hspace{13.2cm}\includegraphics{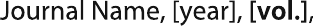}}
\fancyfoot[CE]{\vspace{-7.2pt}\hspace{-14.2cm}\includegraphics{RF}}
\fancyfoot[RO]{\footnotesize{\sffamily{1--\pageref{LastPage} ~\textbar  \hspace{2pt}\thepage}}}
\fancyfoot[LE]{\footnotesize{\sffamily{\thepage~\textbar\hspace{3.45cm} 1--\pageref{LastPage}}}}
\fancyhead{}
\renewcommand{\headrulewidth}{0pt} 
\renewcommand{\footrulewidth}{0pt}
\setlength{\arrayrulewidth}{1pt}
\setlength{\columnsep}{6.5mm}
\setlength\bibsep{1pt}
%%%END OF FOOTER%%%

%%%FIGURE SETUP - please do not change any commands within this section%%%
\makeatletter 
\newlength{\figrulesep} 
\setlength{\figrulesep}{0.5\textfloatsep} 

\newcommand{\topfigrule}{\vspace*{-1pt}% 
\noindent{\color{cream}\rule[-\figrulesep]{\columnwidth}{1.5pt}} }

\newcommand{\botfigrule}{\vspace*{-2pt}% 
\noindent{\color{cream}\rule[\figrulesep]{\columnwidth}{1.5pt}} }

\newcommand{\dblfigrule}{\vspace*{-1pt}% 
\noindent{\color{cream}\rule[-\figrulesep]{\textwidth}{1.5pt}} }

\makeatother
%%%END OF FIGURE SETUP%%%

%%%TITLE, AUTHORS AND ABSTRACT%%%
\twocolumn[
  \begin{@twocolumnfalse}
\vspace{3cm}
\sffamily
\begin{tabular}{m{4.5cm} p{13.5cm} }

\includegraphics{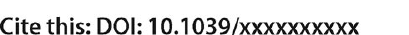} & \noindent\LARGE{\textbf{Self-Diffusion of Glycerol in $\gamma$-Alumina Nanopores: Understanding the Effect of Pore Saturation on the Dynamics of Confined Polyalcohols}} \\%Article title goes here instead of the text "This is the title"
\vspace{0.3cm} & \vspace{0.3cm} \\

 & \noindent\large{Gerardo Campos-Villalobos,\textit{$^{a}$} Flor R. Siperstein,\textit{$^a$} Carmine D'Agostino,\textit{$^a$} and Alessandro Patti$^{\ast}$\textit{$^{a}$}} \\%Author names go here instead of "Full name", etc.

\includegraphics{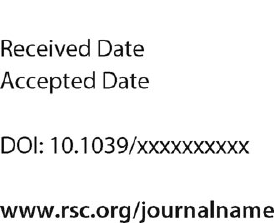} & \noindent\normalsize{Molecular Dynamics simulations of glycerol confined in $\gamma$-Al$_2$O$_3$ slit nanopores are used to explain controversial and inconsistent observations reported in the literature regarding the dynamics of viscous fluids in confined geometries. Analysing the effects of the degree of confinement and pore saturation in this system, we found that the presence of the solid/liquid interface and the liquid/gas interface in partially saturated pores are the main contributors for the disruption of the hydrogen bond network of glycerol. Despite the reduction of hydrogen bonds between glycerol molecules caused by the presence of the solid, glycerol molecules near the solid surface can establish hydrogen bonds with the hydroxyl groups of $\gamma$-Al$_2$O$_3$ that significantly slow-down the dynamics of the confined fluid compared to the bulk liquid. On the other hand, the disruption of the hydrogen bond network caused by the liquid/gas interface in unsaturated pores reduces significantly the number of hydrogen bonds between glycerol molecules and results in a faster dynamics than in the bulk liquid. Therefore, we suggest that the discrepancies reported in the literature are a consequence of measurements carried out under different pore saturation conditions.} \\

\end{tabular}

 \end{@twocolumnfalse} \vspace{0.6cm}

  ]
%%%END OF TITLE, AUTHORS AND ABSTRACT%%%

%%%FONT SETUP - please do not change any commands within this section
\renewcommand*\rmdefault{bch}\normalfont\upshape
\rmfamily
\section*{}
\vspace{-1cm}

%%%FOOTNOTES%%%

\footnotetext{\textit{$^{a}$~School of Chemical Engineering and Analytical Science, The University of Manchester, Sackville Street, M13 9PL, Manchester, UK; E-mail: alessandro.patti@manchester.ac.uk}}

%Please use \dag to cite the ESI in the main text of the article.
%If you article does not have ESI please remove the the \dag symbol from the title and the footnotetext below.
%\footnotetext{\dag~Electronic Supplementary Information (ESI) available: [details of any supplementary information available should be included here]. See DOI: 10.1039/b000000x/}
%additional addresses can be cited as above using the lower-case letters, c, d, e... If all authors are from the same address, no letter is required

%\footnotetext{\ddag~Additional footnotes to the title and authors can be included \textit{e.g.}\ `Present address:' or `These authors contributed equally to this work' as above using the symbols: \ddag, \textsection, and \P. Please place the appropriate symbol next to the author's name and include a \texttt{\textbackslash footnotetext} entry in the the correct place in the list.}

%%%END OF FOOTNOTES%%%

%%%MAIN TEXT%%%%

\section{Introduction}

Spatial confinement in volumes of mesoscopic dimensions is known to dramatically affect the thermodynamic and transport properties of molecular fluids \cite{evans1990}. Changes in the location of first-order transitions, structure and dynamics have been often reported for a wide variety of molecular systems in hard confining geometries \cite{mckenna2005,christenson2001,alba2006,chia2017}. For the specific case of solid-fluid interfaces, the spatial dependence in the intermolecular interactions imposes an inherent inhomogeneity in the system that challenges the scope of classical equilibrium thermodynamics by establishing a density variation perpendicular to the interface \cite{mittal2008}. The fundamental understanding of these modifications from bulk fluid behaviour is of great importance in technologies such as heterogeneous catalysis, engine lubrication, surface coating and enhanced oil recovery. In this regard, statistical mechanical theories and computer simulation techniques, including Monte Carlo (MC) and Molecular Dynamics (MD) methods, have played a crucial role. 

Apart from density functional methods \cite{monson2012}, most theoretical models of confined fluids do not describe spatial inhomogeneities in an explicit way. The phenomenon is rather treated as a phase-equilibrium problem between two macroscopically distinct phases with unique densities: the bulk and confined fluids (\textit{e.g}. Travalloni's model \cite{travalloni2010} and quasi-two dimensional fluid approach \cite{martinez2007}). By contrast, in particle-based simulations the fluid spatial distribution within the confining media is considered explicitly, allowing then for the study of the relationship between molecular structure and mobility.

While a considerable number of computer simulation studies have focused on associating liquids such as water confined in slit-shaped nanopores and nanotubes \cite{mashl2003, matsumoto2002, striolo2005}, only little attention has been paid to the case of organic viscous glass-formers in mesoporous materials, which are of high interest in condensed matter physics. A fundamentally important question in these systems is how the glass transition temperature $T_{\textrm{g}}$ is shifted due to the introduction of two- and three-dimensional geometric constraints \cite{pissis1994}. Within this category, the sugar alcohol glycerol (C$_{3}$O$_{3}$H$_{8}$) has been one of the most extensively studied molecular glass-formers in experiments over the last few decades \cite{debenedetti2001, davidson1951, root1997, lunkenheimer1996}. Despite its apparent simplicity, glycerol is a flexible molecule able to form highly directional intra- and inter-molecular hydrogen bonds. Thus in the liquid state, the complex hydrogen bonding network results in the emergence of an intermediate range order, which is atypical in molecular fluids with only dispersive interactions \cite{chelli1999, blieck2005}. 

Recent experiments on the mobility of liquid glycerol under hard confinement have led to especially intriguing, but apparently contradictory, conclusions \cite{buntkowsky2018}. On the one hand, a breaking of the conventional free volume rules for molecular mobility was reported from spectroscopy measurements of glycerol in a mesoporous silica glass \cite{kilburn2008}. The authors observed that upon confinement, the molecular motion was strongly suppressed, but that the local free volume between molecules was higher than in the bulk phase. On the other hand, D'Agostino et al. \cite{d2012} reported an unexpected significant enhancement in the self-diffusion of glycerol when confined in mesoporous gamma-alumina ($\gamma$-Al$_{2}$O$_{3}$). The ratio of the effective self-diffusivity within the porous media to the free bulk liquid self-diffusivity was found to be $\sim 1.33$ at $20^{\circ}$C and atmospheric pressure. Anomalously high self-diffusion coefficients had previously been observed for water and alkanes in partially filled porous materials \cite{d1989enhanced, valiullin1997}. Such a phenomenon has generally been attributed to the fast interphase exchange between the vapour and condensed phases within the confining volume. Nevertheless, although it is well known that the critical point of the vapour-liquid coexistence in simple fluids is shifted to lower temperatures in confined environments \cite{evans1986, dijkstra1997}, these arguments do not explain the high mobility of glycerol due to its negligible vapour pressure at the studied temperatures. It was hypothesized that the perturbation in the translational molecular motion was generated by a disruption in the hydrogen bond network of the liquid due to topological defects in the solid substrate \cite{d2012}. Previous computer simulation studies of glycerol in fully saturated cylindrical channels of hydroxylated silica have however demonstrated that confinement induces the classical effects seen for simple van der Waals glass-forming liquids, including structural heterogeneities and a dramatic slowing down in the overall relaxation dynamics \cite{busselez2009}.

Inspired by such an unsolved conundrum, we herein investigate the structure and dynamics of glycerol in realistic slit-shaped pores of $\gamma$-Al$_{2}$O$_{3}$ by performing atomistic MD simulations. In particular, we focus our attention on the effects of pore size and liquid concentration. Our results suggest that a necessary condition for the enhancement in the molecular diffusion over translational degrees of freedom is the partial saturation of the pores. In such conditions, the formation of interfaces with vacuum regions is found to profoundly affect the kinetics of breaking and re-formation of hydrogen bonds.\\

%%%%%%%%%%%%%%%%%%%%%%%%%%%%%%%%%%%%%%%%%%%%%%%%%%%%%%%%%%%%%%%%%%%%%

\section{Computational Methods}
\subsection{Models}
Classical MD simulations were performed using atomistic-scale representations of glycerol and $\gamma$-Al$_{2}$O$_{3}$ crystals. Interactions of glycerol were described via inter- and intra-molecular potentials of the OPLS/AA force-field family \cite{jorgensen1996}. In particular, we adopted the refined parameters by Caleman et al. \cite{caleman2011}, which allow for an accurate reproduction of structural and dynamical properties of the condensed phase over a wide range of thermodynamic state points. 

\begin{figure}[h]
\begin{center}
\includegraphics[scale=0.25]{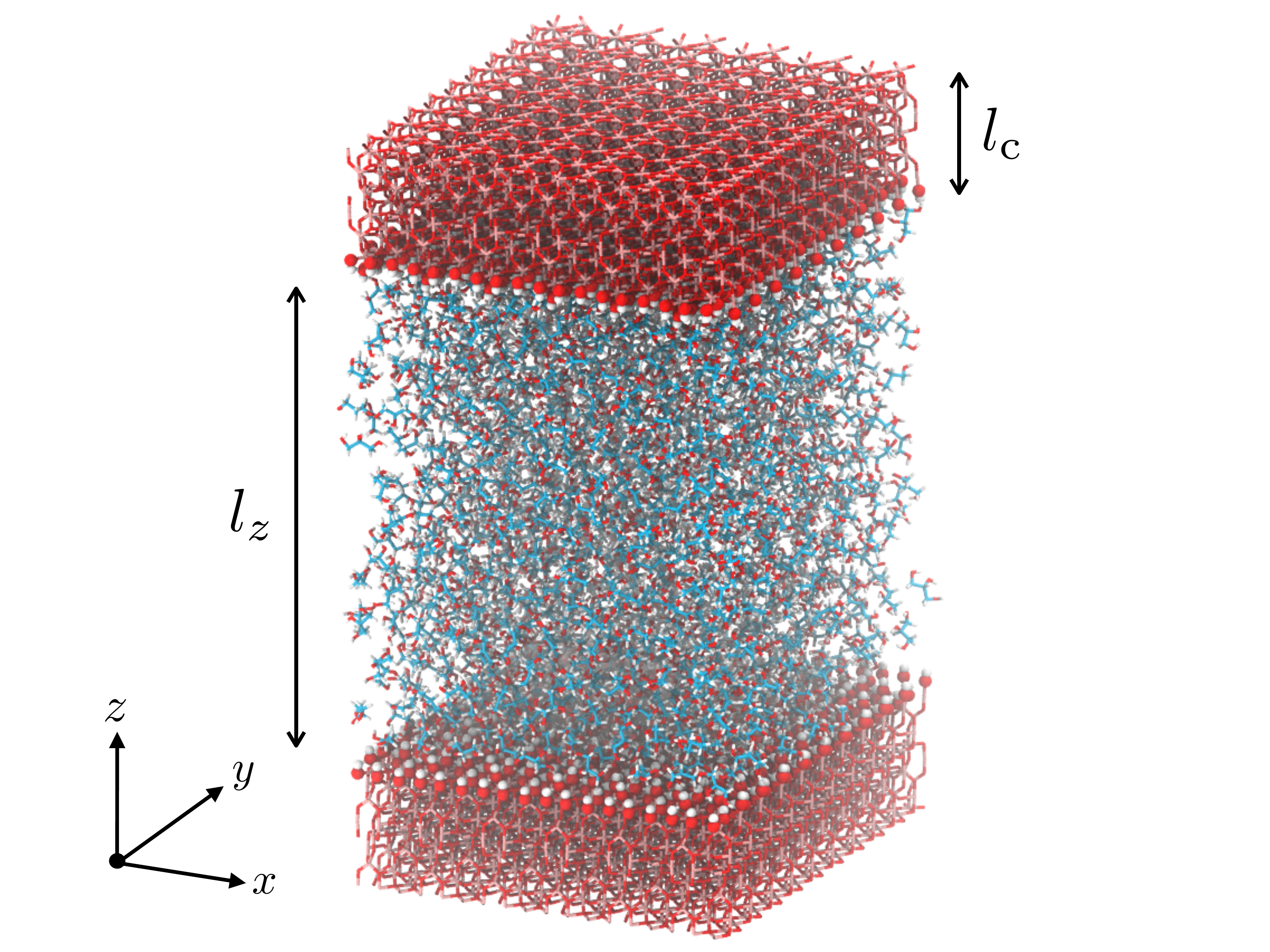}
\caption{Schematic representation of glycerol confined in a slit-shaped pore of $\gamma$-Al$_{2}$O$_{3}$. $l_{\textrm{c}}$ and $l_{z}$ represent the crystal thickness and confinement length, respectively. Cyan, white, red and pink sites indicate carbon, hydrogen, oxygen and aluminum atoms, respectively.}
\label{crystal}
\end{center}
\end{figure}

The crystallographic morphology of bulk $\gamma$-Al$_{2}$O$_{3}$ due to Digne et al. \cite{digne2004} was employed to construct the crystal structure. More specifically, the unit cell, which is comprised of 16 Al atoms and 24 O atoms, was replicated into a relatively large crystal supercell of dimensions $10a\times6b\times2c$, with $a=5.587$  $\textup{\AA}$, $b=8.413$ $\textup{\AA}$, and $c=5.587$ $\textup{\AA}$. This finite crystal size offers a reasonable balance between exposed surface area ($A_{c}=55.870 \times 50.478$ $\textup{\AA}^{2}$) and thickness ($l_{c}=11.174$ $\textup{\AA}$) while maintaining a computationally affordable number of atomic sites. In order to create a realistic surface structure, the crystal was cut along the (100) crystallographic facet, which is one of the main surfaces exposed by $\gamma$-Al$_{2}$O$_{3}$ nanocrystals. The final fully-hydroxylated surface of the crystal model contained $\sim 17$ OH groups per square nanometer with the OH bond vectors aligned in the direction normal to the surface. In order to mimic slit-shaped pore geometries, two identical layers of the solid substrate were placed in front of each other at a mutual distance $l_z$ along the $z$ direction of the simulation box. Different distances between the outermost surface H atoms of specularly symmetric crystalline $\gamma$-alumina layers were considered: $l_{z}=$ 20, 40 and 60 $\textup{\AA}$. These inter-wall distances lie within the range of the typical pore dimensions of $\gamma$-Al$_{2}$O$_{3}$ nanocrystals \cite{d2012}. Since the systems were simulated under three-dimensional periodic boundary conditions, an extra empty gap of thickness $\sim 200$ $\textup{\AA}$ was considered between the non-hydroxylated faces of the crystals in order to prevent long-range electrostatic interactions between periodic images of the primary slab. A pictorial representation of the model system is presented in Fig.\ \ref{crystal}.\\

Crystal atoms were treated as stationary (rigid) charged Lennard-Jones (LJ) sites with parameters taken from the CLAYFF force-field \cite{cygan2004}. By effectively freezing the crystal atoms, we make the following assumptions: (i) the rotation and translation of the crystals is negligible within the timeframe of a simulation run; and (ii) the orientation in which we freeze the crystal surface atoms plays a minor role in determining the properties of the interfacial glycerol molecules. LJ parameters for unlike interactions between glycerol and $\gamma$-Al$_{2}$O$_{3}$ were determined via geometric mixing rules as prescribed by Striolo and coworkers \cite{phan2013}. 

Although the classical atomistic model adopted here allows for the study of complex liquid-solid interactions, such as hydrogen-bonding, it neglects possible changes of the bonding nature, including chemisorption events. Nevertheless, a more detailed description based on quantum-mechanical calculations is impractical considering the system size and time-scale investigated in the present study. Other interesting alternatives such as the ReaxFF methodology \cite{van2001}, which approximates chemical reactivity with a classical treatment, remain computationally expensive.

\subsection{Simulation Details}
Glycerol molecules were initially distributed randomly between the two fully-hydroxylated $\gamma$-Al$_{2}$O$_{3}$ crystalline layers, so as to retrieve the equilibrium density of the bulk fluid, $\rho_{\textrm{bulk}}=1.2$ g cm$^{-3}$, obtained from isothermal-isobaric simulations of the bulk liquid phase at $T=298$ K and $P=1$ bar. Additionally, in order to study the effect of nanopore saturation at each confinement length $l_{z}$, different confinement concentrations, $\rho_{\textrm{conf}}$, have been investigated. The confinement concentration, which can be regarded as the liquid concentration in the slit-shaped pore, is calculated as
\begin{equation}
\rho_{\textrm{conf}} = \frac{NM_{\textrm{w}} N_{\textrm{A}}  }{A_{c}  l_{z}}
\end{equation}
where $N$, $M_{\textrm{w}}$ and $N_{\textrm{A}}$ are the number of glycerol molecules, glycerol molecular weight and Avogadro's number, respectively. In particular, we studied systems with $\rho_{\textrm{conf}} = \left[ 0.4, 0.6, 0.8, 1.0, 1.2\right]$ g cm$^{-3}$. The number of glycerol molecules in each simulation is given in Table \ref{systems}.

IIn the first stage, high-energy structures were relaxed by applying the steepest descent method. Subsequently, MD simulations of the confined systems under three-dimensional periodic boundary conditions were performed in the canonical ensemble ($NVT$) at $T=300$ K using the GROMACS 5.0.4 package \cite{van2005}. The classical equations of motion were numerically integrated using the leapfrog algorithm with a time step of 2 fs. The Nos\'{e}-Hoover thermostat \cite{nose1984,hoover1985} with a relaxation constant of 0.5 ps was employed in order to control temperature fluctuations. The global cut-off for non-bonded (dispersion and electrostatic) interactions was set to the OPLS/AA value of $11$ $\textup{\AA}$. Due to the slab geometry with clearly defined interfaces, standard analytical tail corrections to the energy and pressure for bulk systems were neglected. Long-range electrostatics was handled using the particle mesh Ewald (PME) method \cite{darden1993}. For efficiency purposes, all bond-lengths in the glycerol molecules were constrained to their equilibrium values using the LINCS algorithm \cite{hess1997}, applying two iterations for a correct energy conservation. Structural and dynamical properties were calculated from long trajectories of a duration of 100 ns containing $\sim 5\times10^{4}$ thermalized configurations.\\

\begin{table}[h]
 \centering
  \caption{Number of glycerol molecules ($N$) at different confinement concentrations ($\rho_{\textrm{conf}}$) in pores of confinement length $l_{z}$. Eventual occurrence of a vacuum-liquid interface (VLI) is also specified.}
  \label{systems}
  \begin{tabular}{l c c c}
    \hline    
    \hline
    $l_{z} (\textup{\AA})$ & $N$ & $\rho_{\textrm{conf}}$ (g cm$^{-3}$) & VLI Formation \\
    \hline    
        60& 442& 0.4 &Yes\\
        & 663& 0.6 &Yes\\
        & 885& 0.8 &Yes (bubble)\\
       & 1106& 1.0 &Yes (bubble)\\
       & 1394& 1.2 &No\\
      & & & \\
    40& 295& 0.4 &Yes\\
        & 442& 0.6 &Yes\\
        & 590& 0.8 &Yes (bubble)\\
       & 738& 1.0 &Yes (bubble)\\
       & 930& 1.2 &No\\
      & & & \\
    20& 147& 0.4  &Yes\\
        & 221& 0.6  &Yes\\
        & 295& 0.8  &Yes\\
       & 369& 1.0  &Yes (bubble)\\
       & 464& 1.2  &No\\
    \hline
    \hline
  \end{tabular}
\end{table}

\section{Results }

\begin{figure}[h]
\begin{center}
\includegraphics[scale=0.23]{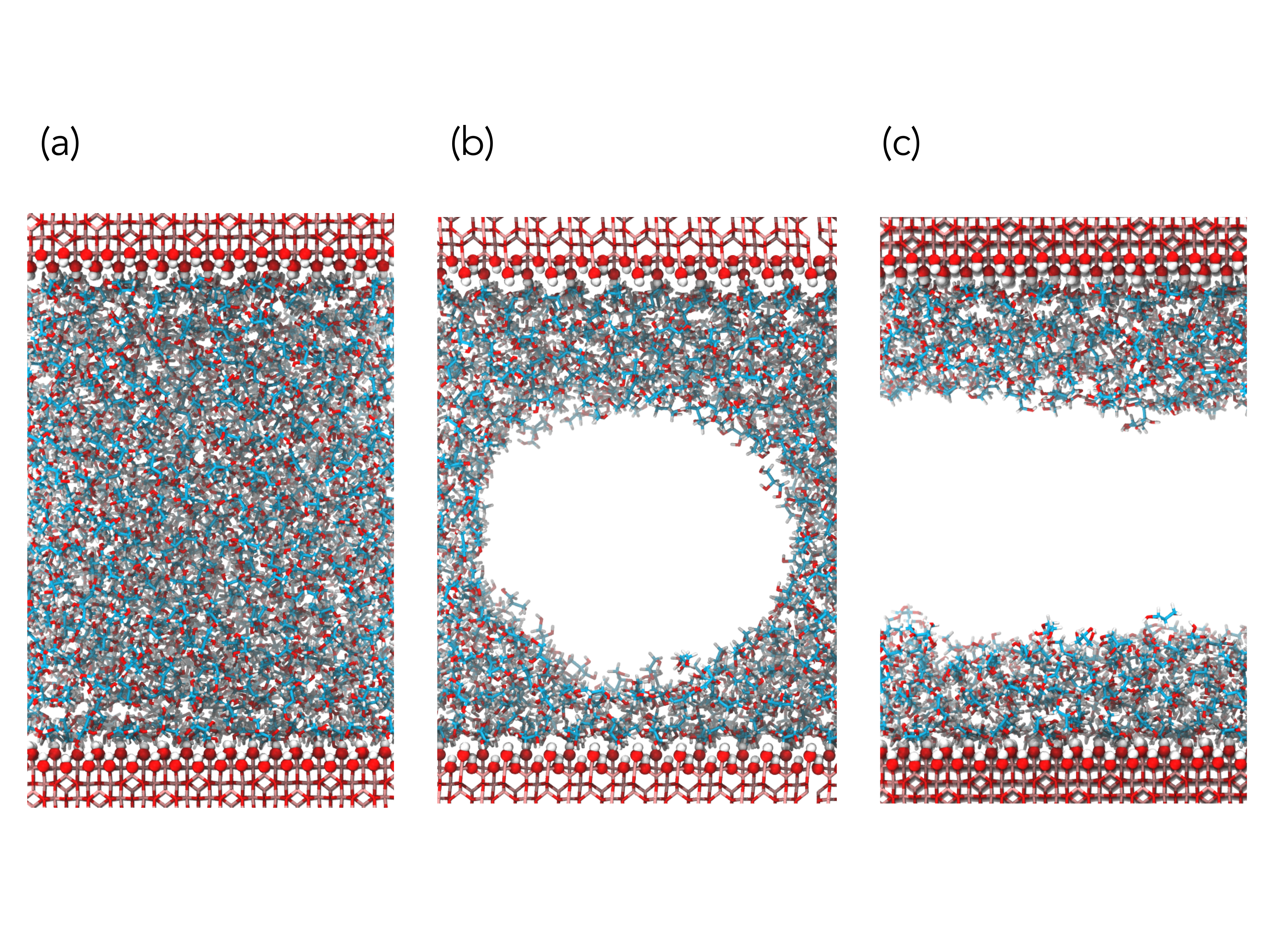}
\caption{Glycerol in $\gamma$-Al$_2$O$_3$ slit nanopores at confinement concentrations $\rho_{\text{conf}} = 1.2$ (a), 0.8 (b) and 0.6 (c) g cm$^{-3}$ at $l_z=60$ $\textup{\AA}$. Bubbles within the fluid (b) and fully separated films (c) are generally formed for $0.8 \le \rho_{\textrm{conf}} \le 1.0$ g cm$^{-3}$ and $\rho_{\textrm{conf}} \le 0.6$ g cm$^{-3}$, respectively.}
\label{bubbles}
\end{center}
\end{figure}

In this section, we show that hydrogen-bonding networks play an important role in the translational dynamics of glycerol and are influenced by confinement. However, we will see that the enhanced mobility of glycerol under confinement is not simply the consequence of the disruption of its intermolecular hydrogen-bonding network resulting from the presence of the pore walls as suggested in Ref.\ \cite{d2012}. Our results unambiguously indicate that the interactions established at the solid-liquid interface and the degree of confinement play both a key role in determining the glycerol dynamics in $\gamma$-Al$_2$O$_3$ nanopores. In particular, the degree of confinement depends on the pore size and on the glycerol concentration in the pore. The former is quantified in terms of distance, $l_z$, between two $\gamma$-Al$_2$O$_3$ surfaces facing each other as shown in Fig.\ \ref{crystal}; whereas the latter, $\rho_\text{conf}$, is the mass of glycerol in the volume available between these solid surfaces (see Eq.\ 1) and measures the level of pore saturation. In Fig.\ \ref{bubbles}, one can appreciate the effect of reducing  $\rho_\text{conf}$ on the distribution of glycerol in the pore for the specific case of $l_z=60$  $\textup{\AA}$. Upon decreasing the relative confinement concentration $\rho_\text{conf}/\rho_\text{bulk}$ from 1 to 1/2 (left to right frame in Fig.\ \ref{bubbles}), we detect the nucleation of a low-density cavity (or bubble), which eventually leads to two fully separated liquid films adsorbed at the crystal surfaces and entrapped between a solid-liquid and a gas-liquid interface. Given the low vapour pressure of glycerol, no molecules were detected in the gas phase, which for practical purposes can be effectively considered as vacuum.

\begin{figure}[h]
\begin{center}
\includegraphics[scale=0.33]{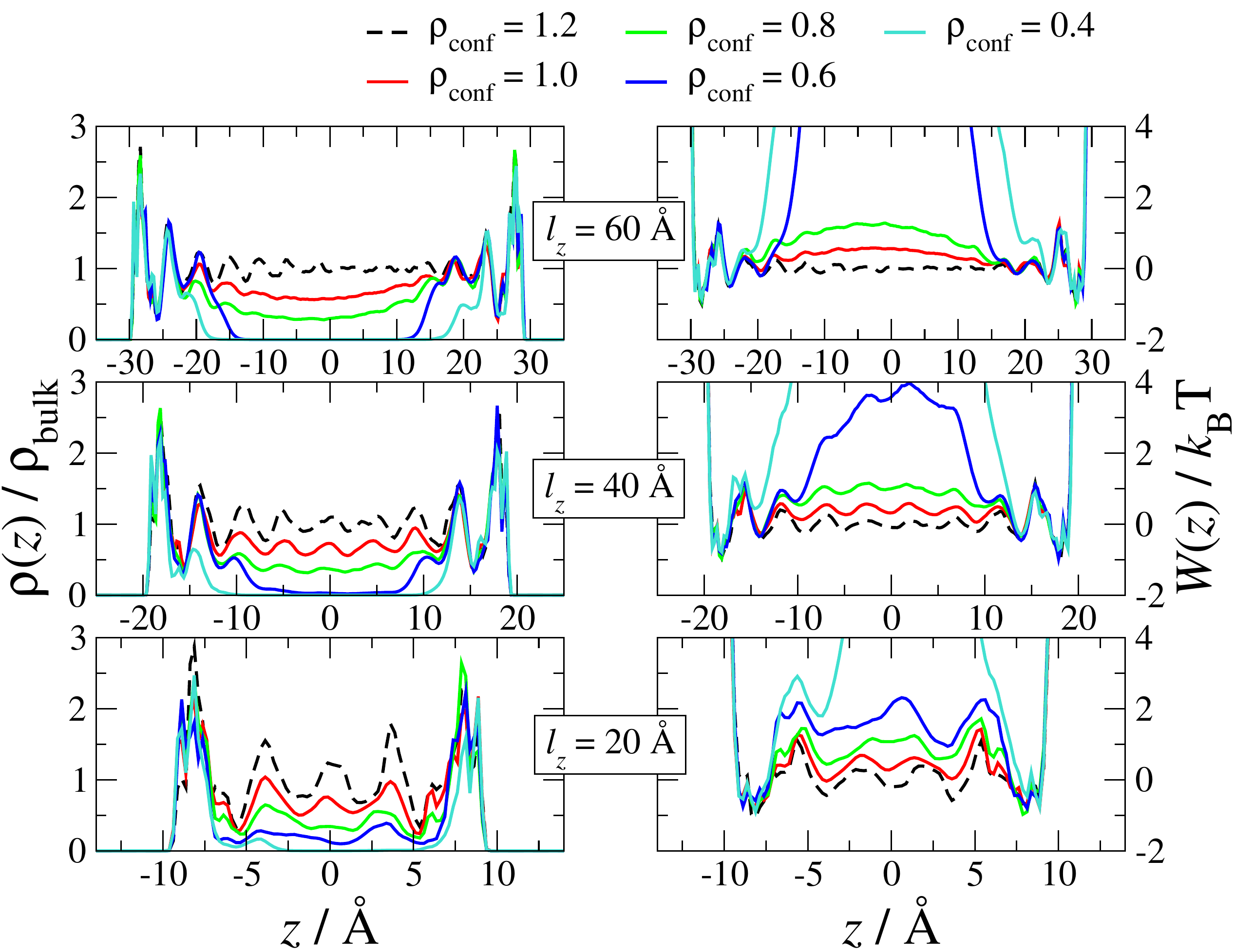}
\caption{Density profiles (left) and free energy profiles (right) of glycerol along the $z$ coordinate in slit nanopores of different confinement length $l_{z}$. All the confinement concentrations for each case are included. $\rho(z)$ is normalized by the bulk density $\rho_{\text{bulk}}$ and $\mathcal{W}(z)$ by $k_{\text{B}}T$. }
\label{dens_all}
\end{center}
\end{figure}

The value of the confinement concentration at which bubbles and separated films are observed, strictly depends on the pore size or, equivalently, on the confinement length $l_z$, as reported in Table 1. Both $\rho_\text{conf}$ and $l_z$ have a remarkable impact on the distribution of glycerol in the pore and on its ability to form structural correlations over the length scales imposed by the degree of confinement. To gain a more detailed insight into the effect of the degree of confinement, in Fig.\ \ref{dens_all} we present the glycerol mean local density profiles along the direction $z$ perpendicular to the solid surfaces at  $1/3 \le \rho_\text{conf}/\rho_\text{bulk} \le 1$ and $20 \le l_z \le 60$ $\textup{\AA}$. A fully saturated pore, where $\rho_\text{conf}=\rho_\text{bulk}$, is characterized, in the vicinity of the solid-liquid interface, by strong spatial correlations that are propagated, less and less intensively, to the glycerol molecules approximately located in the center of the pore, where $z=0$, relatively far from the solid surface. In this case, no bubbles or vacuum-liquid interfaces are observed.  However, upon decreasing the glycerol confinement concentration, it is not possible to fully permeate the pore, which remains unsaturated, and a low-density cavity starts to form. The threshold value of the confinement concentration, $\rho_\text{conf}^*$, at which the formation of such a vacuum domain is observed, is determined by the pore size. In general, at $\rho_\text{conf}/\rho_\text{bulk} = 0.5$ all pores exhibit two fully separated liquid films and a clear vacuum region, regardless of their size. 

An interesting feature observed at all the confinement lengths studied here, is that changes in the liquid concentration barely affect the shape of the density distributions close to the surface. Taking as a reference the case of $l_{z}=60$ $\textup{\AA}$, the first two large peaks in $\rho(z)$ remain effectively unaltered in the whole range of pore saturations. This indicates that even at partial saturation, the correlations of glycerol with $\gamma$-Al$_2$O$_3$ surface atoms are still very significant. Based on the suppression of the oscillations in the density profiles, the influence of the substrate can be considered to be effective on glycerol molecules lying within a distance of approximately $20$ $\textup{\AA}$ from the solid surface. 

The strong spatial dependency observed in the molecular packing of the fluid is the result of the local free energy, $\mathcal{W}(z)$, of transferring a glycerol molecule from the bulk phase to a slab centred at position $z$ \cite{marrink1994}:

\begin{equation}
\frac{\mathcal{W}(z)}{k_BT} = - \ln \frac{\rho(z)}{\rho_{\text{bulk}}}
\end{equation}
where $k_{\text{B}}$ is the Boltzmann's constant. The equilibrium local density distribution, $\rho(z)$, is measured by the probability of finding a glycerol molecule, with center of mass in $z_{0}$, in the interval ($z$, $z+dz$). This probability is proportional to a \textit{constrained} partition function, $Q'(z)$, which  reads 

\begin{equation}
Q'(z) = \mathcal{C} \int { \delta \left( z_{0} - z \right) \exp{\left[ -\beta \mathcal{H}\left( \boldsymbol{r}^{N}, \boldsymbol{p}^{N} \right)\right]} d\boldsymbol{r}^{N}d\boldsymbol{p}^{N} }.
\end{equation}
In particular, $Q'(z) \propto \rho(z)$. In Fig.\ \ref{dens_all}, we report $\mathcal{W}(z)$ for all the systems studied here. The presence of significant energy barriers in $\mathcal{W}(z)$ in the vicinity of the crystal suggests that glycerol molecules are highly constrained in specific regions of space and rarely diffuse to contiguous layers at the studied temperature. The first minima, which exceeds the mean attractive energy in the bulk fluid by approximately 1 $k_{\text{B}}T$ reflects the favourable interactions established with the solid. This explains in part, the formation of the fully separated adsorbed liquid films at $\rho_{\text{conf}} \leq 0.6$ g cm$^{-3}$.

\begin{figure}[h]
\begin{center}
\includegraphics[scale=0.18]{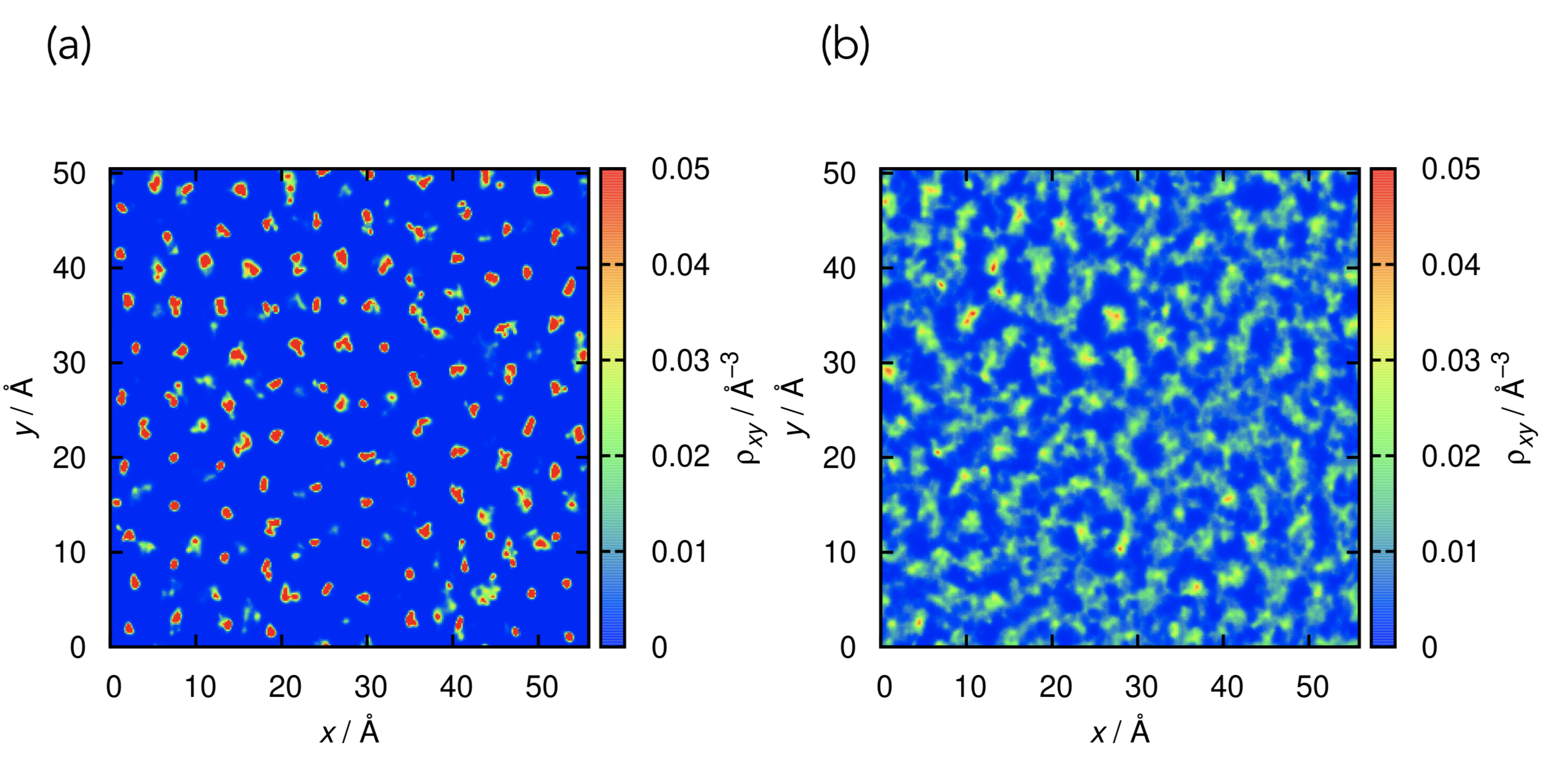}
\caption{Two-dimensional density distributions of the glycerol's central carbon atoms, C$_{\textrm{c}}$, in the adsorbed layer (a) and in the \textit{unperturbed} bulk fluid (b), for $l_z=60$ $\textup{\AA}$.}
\label{maps}
\end{center}
\end{figure}

The solid-liquid spatial correlation can also be appreciated from the two-dimensional probability density maps of the glycerol's central carbon, $C_c$, reported in Fig.\ \ref{maps}. In particular, the left frame refers to the layer of glycerol molecules adsorbed on the solid surface and indicates the presence of specific preferential locations that these molecules are more likely to occupy. Therefore, not only is a partial, but no-negligible, ordering found in the direction $z$ perpendicular to the interface, but also in the plane components ($x,y$) parallel to it. Similarly to the perpendicular spatial correlations in the liquid, also the in-plane spatial correlations become very weak closer to the center of the pore (right frame of Fig.\ \ref{maps}). The strongly localized molecular positions of glycerol found in the adsorbed layer clearly indicate the presence of energetic interactions able to overcome the penalty due to the loss of configurational entropy. We attribute this in part to the propensity of glycerol to form highly-directional hydrogen bonds with the $\gamma$-Al$_2$O$_3$ hydroxyl groups.

\begin{figure}[h]
\begin{center}
\includegraphics[scale=0.33]{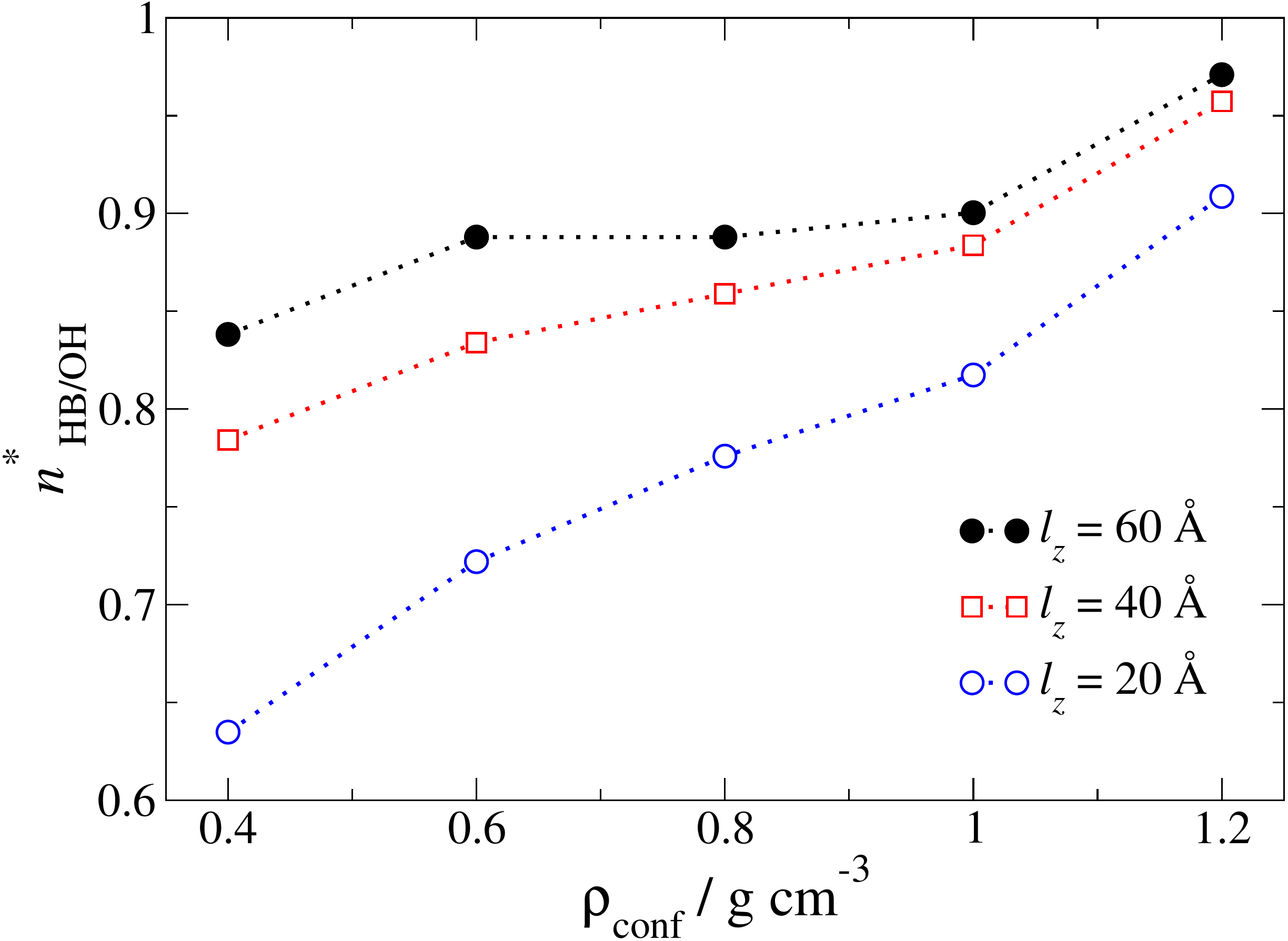}
\caption{Average number of hydrogen bonds per OH group in glycerol as a function of the confinement concentration, $\rho_{\text{conf}}$ in slit nanopores of different confinement length $l_{z}$. The results are normalized by the value in the bulk $\left<  n_{\textrm{HB/OH,bulk}} \right>$. Dashed lines are a guide for the eye.}
\label{nhb_ratios}
\end{center}
\end{figure}

With such a detailed insight into the spatial arrangement of glycerol in $\gamma$-Al$_2$O$_3$ nanopores, we now investigate how the glycerol hydrogen-bond network is being affected by the formation of solid-liquid and vacuum-liquid interfaces. To this end, we first need to introduce a geometric criterion defining the conditions for the formation of a hydrogen-bond. Following the suggestion by Busselez and coworkers, who studied the dynamics of glycerol in cylindrical silica nanopores, two hydroxyl groups form a hydrogen bond if the distance between their oxygen atoms is smaller than 3.5 $\textup{\AA}$ and simultaneously, the angle between two hydroxyl groups  O$-$H$\cdots$O $\leq120^{\circ}$ \cite{busselez2009}. In agreement with these authors' MD simulation results, we find that the average number of hydrogen bonds per OH group in the unconstrained liquid phase is $\left<  n_{\textrm{HB/OH,bulk}} \right> \approx 2$. This value changes if glycerol is confined in $\gamma$-Al$_2$O$_3$ as reported in Fig.\ \ref{nhb_ratios}, where the reduced number of hydrogen bonds per OH group $n^*_\text{HB/OH} \equiv \left<  n_{\text{HB/OH,conf}} \right> / \left< n_{\text{HB/OH,bulk}} \right>$ is reported as a function of $\rho_\text{conf}$ for different pore sizes.

The trend observed in Fig.\ \ref{nhb_ratios} unveils the remarkable effect of the degree of confinement on the disruption of the the hydrogen-bonding network of glycerol. At $\rho_{\text{conf}}=\rho_{\text{bulk}}$ such a disruption is not especially significant, with $n^*_\text{HB/OH} > 0.9$. In this case, the reduction in $n^*_\text{HB/OH}$ is caused exclusively by the presence of the $\gamma$-Al$_2$O$_3$ surfaces. Although this result conciliates with the suggestion by D'Agostino et al. \cite{d2012}, the reduction of glycerol-glycerol hydrogen bonds is partially compensated by the presence of hydrogen bonds between glycerol and the hydroxyl groups in the surface of $\gamma$-Al$_2$O$_3$. In fact, it was observed that all the glycerol molecules lying in the contact layer have their three OH groups pointing to the surface and forming hydrogen bonds with the $\gamma$-Al$_2$O$_3$ hydroxyl groups. Despite the orientational restrictions imposed by the solid surface, it is important to note that the number of glycerol-glycerol hydrogen-bonds would be smaller if there were no glycerol-glycerol hydrogen-bonds in the contact layer as this would lead to approximately $n^*_\text{HB/OH} \approx 0.85$, $0.77$, and $0.54$ for the pores of $60$ , $40$ and $20$ $\textup{\AA}$ respectively. Therefore, some of the molecules in the contact layer are forming hydrogen-bonds with other glycerol molecules.

The disruption becomes more relevant at lower confinement concentrations, especially if the confinement length is particularly small ($l_z = 20$ $\textup{\AA}$). In other words, if the pore is not fully saturated, the average number of hydrogen bonds per OH group in liquid glycerol significantly decreases, due to the formation of a low-density cavity or two separated liquid films that create a vacuum-liquid interface. The minimum in $n^*_\text{HB/OH}$ is found at the lowest glycerol concentration due to the large interfacial area between the liquid films and vacuum.

\begin{figure}[h]
\begin{center}
\includegraphics[scale=0.33]{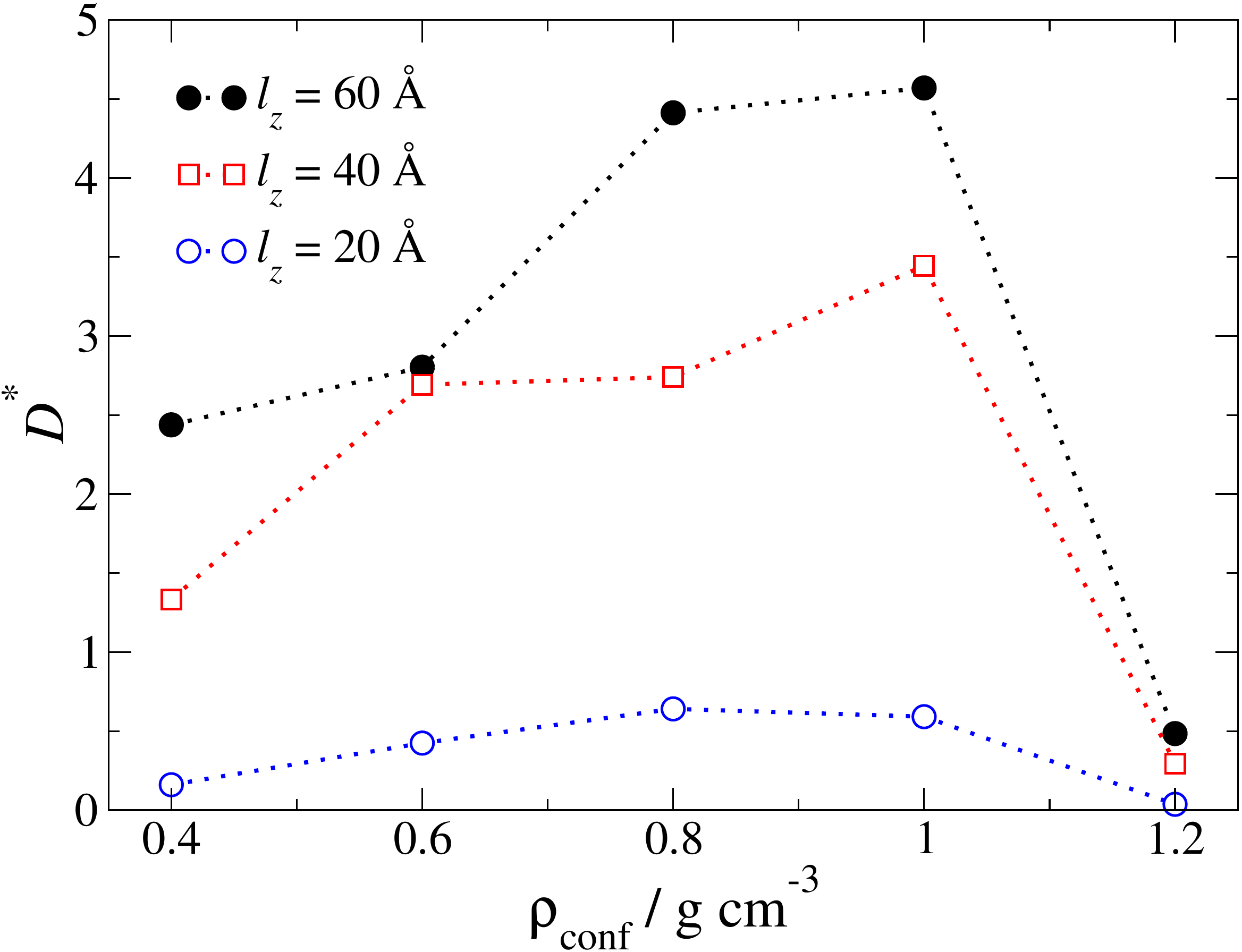}
\caption{Average self-diffusion coefficient of glycerol in slit nanopores of size $l_{z}$ at different confinement concentrations $\rho _{\textrm{conf}}$. The values are normalized with the diffusion coefficient in the isotropic bulk fluid $D_{\textrm{bulk}}$. Dashed lines are added as a guide for the eye.}
\label{dif_ratios}
\end{center}
\end{figure}

Thus, if there is an effect of the hydrogen-bond network on the dynamics of glycerol, then it must be reflected in the dependence of the glycerol diffusivity on the degree of pore saturation. To test the validity of this hypothesis, we have calculated the reduced self-diffusion coefficient of glycerol, $D^* \equiv D_\text{conf}/D_\text{bulk}$, defined as the ratio between the self-diffusivity in the pore and the self-diffusivity in the unconstrained liquid bulk. However, due to the intrinsic anisotropy and spatial inhomogeneity in the confined fluid, neither the standard three-dimensional Einstein relation nor the Kubo relation over the whole confined volume would be suitable approaches \cite{liu2004}. Therefore, $D_\text{conf}$ has been calculated as the density-weighted average of the local parallel diffusion coefficient in finite slabs centered at different $z$ positions in the simulation box:
\begin{equation}
D_\text{conf} \equiv \left< D_{||}(z)\right> = \frac{\int_{l_{z}}{D_{||}(z) \rho (z) dz} }{\int_{l_{z}}{\rho(z) dz}}
\end{equation}
In particular, the two-dimensional parallel component of the diffusion coefficient in a given slab $\delta_z$, defined by the spatial interval ($z$,$z+dz$), reads 
\begin{equation}
D_{||}\left( \delta_z\right) = \lim_{\tau \rightarrow \infty}{\frac{\left< |\boldsymbol{r}_{xy}\left(t + \tau \right) - \boldsymbol{r}_{xy} \left( t\right) |^{2} \right>_{\delta_z}}{4\tau P_{\delta_z}\left( \tau \right)}}
\end{equation}
where $\boldsymbol{r}_{xy}(t)$ is the two-dimensional position vector of a glycerol molecule in slab $\delta_z$ at time $t$ and $P_{\delta_z}\left( \tau\right)$ is the survival probability for particles to remain in that slab (see Ref. \cite{liu2004}). The brackets stand for an ensemble average over sample molecules and time origins $t$. Earlier computational studies have demonstrated that the diffusion coefficient calculated by averaging the translational motion along the $z$ direction, $D_\perp$, would provide the same qualitative results \cite{liu2004, mittal2008}. Consequently, even if the global effective three-dimensional diffusion coefficients could be somehow obtained without a decoupling in the translational degrees of freedom, they would (in principle) follow the same trend and be quantitatively similar to those obtained with Eq. 4. The resulting dependence of $D^*$ on the confinement concentration is reported in Fig.\ \ref{dif_ratios} for different confinement lengths.

\begin{figure}[h]
\begin{center}
\includegraphics[scale=0.33]{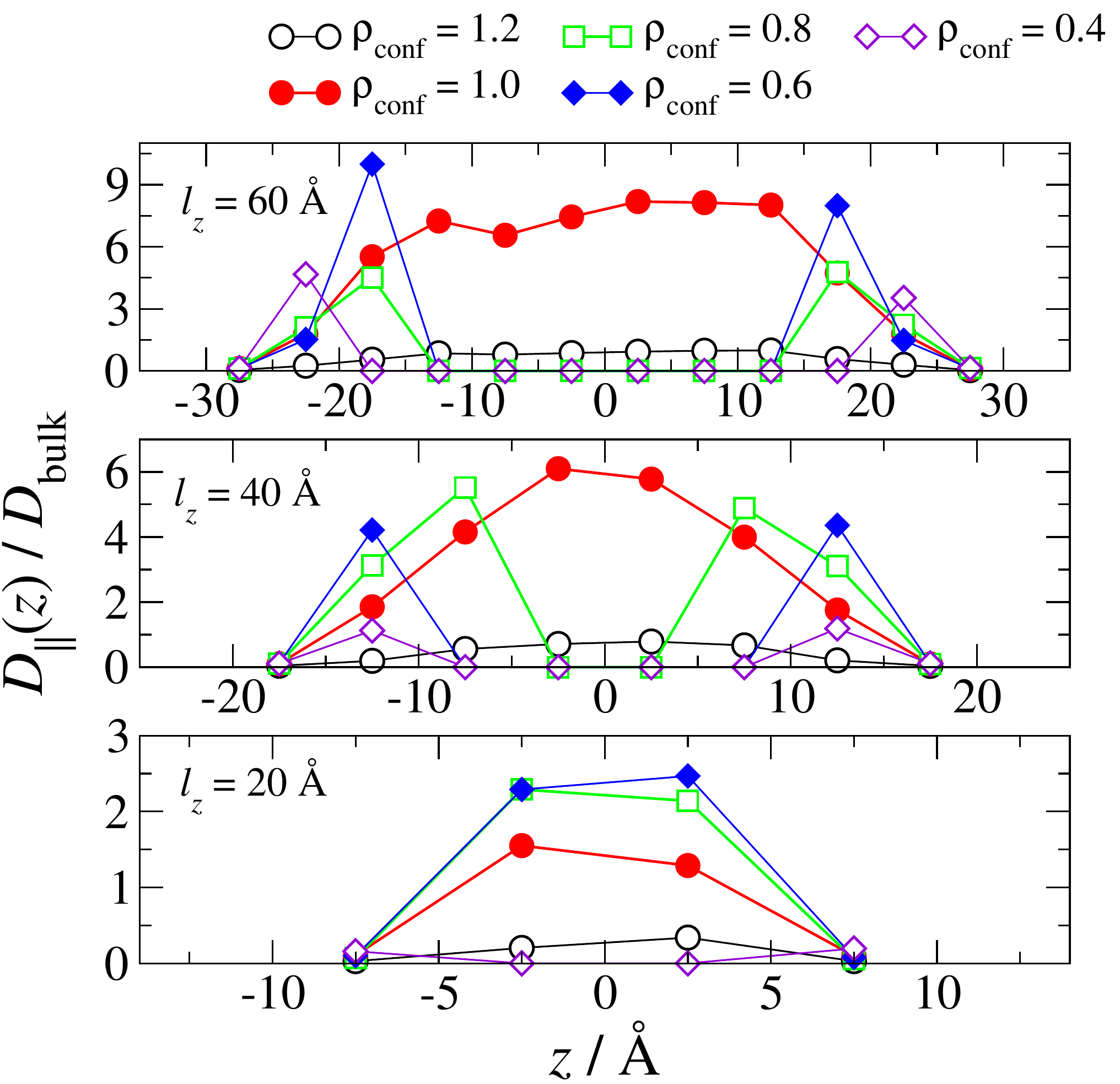}
\caption{Parallel component of the reduced self-diffusion coefficient as a function of the $z$ position for glycerol in slit $\gamma$-alumina nanopores for different $l_{z}$ and confinement concentrations. The local values are normalized by the diffusion coefficient in the isotropic bulk fluid $D_{\textrm{bulk}}$. Solid lines are added as a guide for the eye.}
\label{dif_latz}
\end{center}
\end{figure}

It is interesting to note that at $\rho_\text{conf}/\rho_\text{bulk}=1$ (pore completely saturated), $D^*<1$ for any $l_z$, suggesting a slower diffusion in the pores than in the bulk. By contrast, in pores that are partially saturated ($\rho_\text{conf}/\rho_\text{bulk}<1$), the reduced self-diffusion coefficient increases significantly, especially so at $l_z \ge 40$ $\textup{\AA}$. As soon as a fully developed vacuum-liquid interface forms, $D^*$ decreases again, indicating that the disruption of hydrogen bonds in the liquid glycerol cannot be the only factor determining its dynamics. To unveil the missing piece of this puzzle, one should consider the spatial correlations discussed above and reported in Figs.\ \ref{dens_all} and \ref{maps}. These correlations determine the diffusion profile along the $z$ direction that is shown in Fig.\ \ref{dif_latz}, where we report the local translational diffusion coefficients obtained via Eq 5.

Each frame refers to a different pore size and, within each frame, we report the dependence of the parallel self-diffusivity on the position for different degrees of pore saturation. One can appreciate that the self-diffusion coefficient at the solid-liquid interface is especially low, with $D_{||}(z)/D_\text{bulk}<<1$, but it increases significantly at increasing distance from the solid support. Additionally, at relatively low confinement concentrations, when a vacuum-liquid interface forms, no glycerol molecules are found within the vacuum region and therefore the self-diffusion coefficients are practically meaningless there. Pores that are completely filled show a similar behaviour, but the glycerol self-diffusion coefficient in there remains lower than that in the bulk across the whole pore volume. The case of very small pores ($l_z=20$ $\textup{\AA}$) is especially interesting. While $D^*$ in Fig.\ \ref{dif_ratios} would indicate a reduced mobility as compared to that in the bulk, a deeper analysis reveals that this strictly depends on how far the glycerol molecules are from the $\gamma$-Al$_2$O$_3$ support (Fig.\ \ref{dif_latz}). The nucleation of the low-density cavity and subsequent formation of a vacuum-liquid interface deeply determine the disruption of the hydrogen-bond network and a significant increase of the glycerol mobility, which we do not observe in fully saturated pores, where a vacuum-liquid interface is not present.

\begin{figure}[h]
\begin{center}
\includegraphics[scale=0.33]{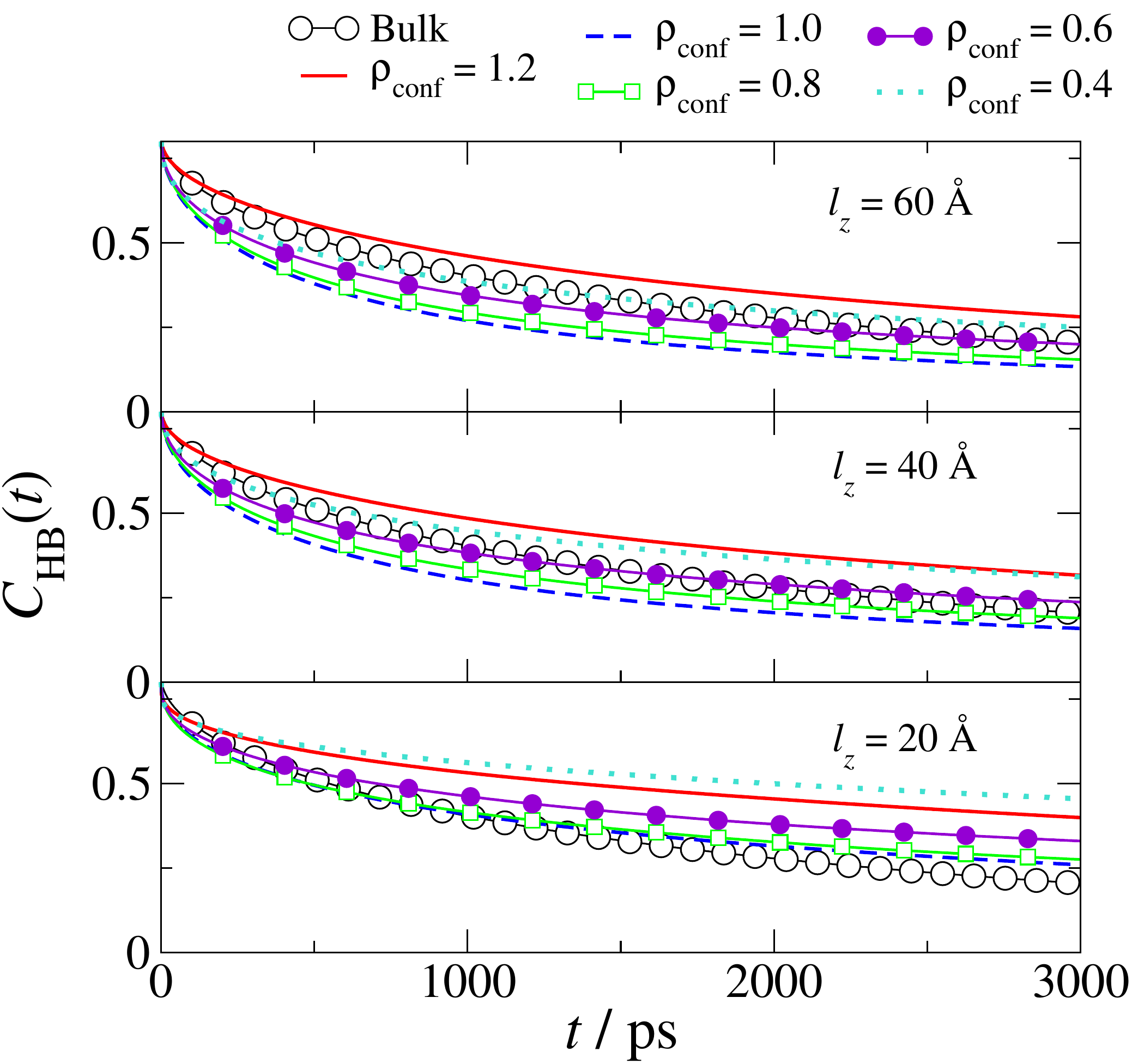}
\caption{Intermittent hydrogen-bonding correlation function in glycerol in the bulk state and under confinement.}
\label{hb_cor}
\end{center}
\end{figure}

Finally, to better understand the effects of the interfaces on the glycerol dynamics, we have assessed the time fluctuations in the hydrogen-bonding network by evaluating the intermittent hydrogen-bond correlation function, $\mathcal{C}_\text{HB}$, firstly introduced by Rapaport in the 1980s \cite{rapaport1983}:

\begin{equation}
\mathcal{C}_{\text{HB}}\left(\tau\right) = \frac{\left< h_{ij}\left( t + \tau\right) \cdot h_{ij} \left(t\right)\right>}{\left< h_{ij} \right>}
\end{equation}
where $h_{ij}\left(t\right)$ is the binary hydrogen bond population operator, which equals unity if the hydroxyl groups $i$ and $j$ are hydrogen-bonded at time $t$ and zero otherwise. In particular, $\mathcal{C}_{\textrm{HB}}\left(\tau\right)$ represents the conditional probability that the hydrogen bond between hydroxyls $i$ and $j$, observed at time $t$, still exists at time $t+\tau$, regardless of whether bond-breaking events might have occurred meanwhile. Basically, a fast decay of $\mathcal{C}_{\textrm{HB}}$ would indicate a relatively short average life-time of a hydrogen bond. In Fig.\ \ref{hb_cor}, we present the correlation's decay for different pore sizes at all confinement concentrations. For comparison, we also include the case of unconstrained bulk liquid. All the curves exhibit a similar time dependence, which can be approximated by an exponential decay. In addition, at $l_{z} \ge 40$ $\textup{\AA}$ and $\rho_{\text{conf}}=1.0$ g cm$^{-3}$, which provide the highest self-diffusivities in Fig.\ \ref{nhb_ratios}, $\mathcal{C}_{\text{HB}}\left(\tau\right)$ displays a faster decays as compared to any other confinement concentration, including the case of the bulk fluid. By contrast, in smaller pores, the correlation decays slower than in the bulk at all times, confirming the above observations. We also calculated the correlation function for the glycerol-surface hydrogen bond kinetics and we found that the curves do not decay to zero in the whole simulated time. This conciliates with the observed suppression in the diffusion of glycerol adsorbed in the crystal surface (Fig. \ \ref{dif_latz}).

\begin{figure}[h]
\begin{center}
\includegraphics[scale=0.33]{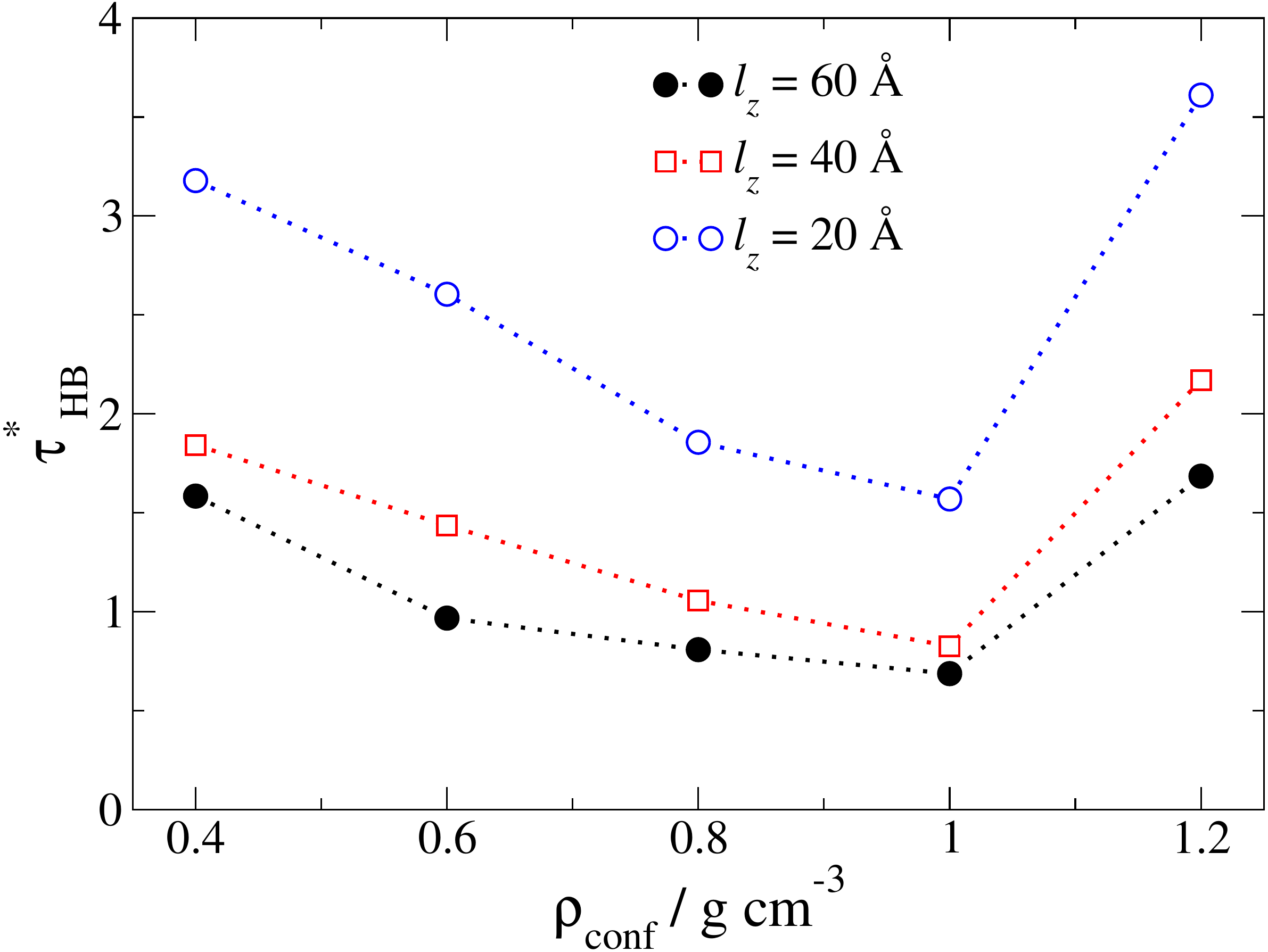}
\caption{Lifetime of the hydrogen bonds between pairs of glycerol molecules in slits of size $l_{z}$ at confinement density $\rho_{\textrm{conf}}$. The values are normalized with the lifetime in the isotropic bulk fluid, $\tau_{\textrm{HB,bulk}}$. Dashed lines are added as a guide for the eye.}
\label{tau_ratios}
\end{center}
\end{figure}

Finally, the average hydrogen bond lifetime, $\tau_{\textrm{HB}}$, was estimated from the reactive flux hydrogen-bond correlation function, $k(\tau)$, which is the time derivative of $\mathcal{C}_{\textrm{HB}}\left(\tau\right)$:

\begin{equation}
k(\tau) = - \frac{d\mathcal{C}_{\text{HB}}\left(\tau\right)}{d\tau}
\end{equation}
and by treating the hydrogen-bond breaking and re-formation as a reversible chemical reaction with well-defined rate constants \cite{luzar1996nature}. The dependence of $\tau_{\text{HB}}$, being normalized by the lifetime in the bulk, on the confinement concentration is reported in Fig.\ \ref{tau_ratios} at different confinement lengths. Overall, the results agree with those of Figs.\ \ref{dif_ratios} and \ref{hb_cor}. As a matter of fact, we note that for glycerol confined in pores with $l_{z} \ge 40$ $\textup{\AA}$, the minimum in the ratio $\tau^{*}_{\text{HB}} = \tau_{\text{HB}}/\tau_{\text{HB,bulk}}$ occurs at the $\rho_{\text{conf}}$ where $D^*$ reaches its maximum value.

\section{Conclusions}
We have investigated the structure and dynamics of glycerol confined in $\gamma$-Al$_{2}$O$_{3}$ slit nanopores using atomistic MD simulations. As observed for simple liquids, the confinement of glycerol within hard geometric boundaries results in a spatially inhomogeneous molecular distribution. More specifically, in the vicinity of the solid surface, changes from a liquid-like to a crystal-like structure are promoted, due in part to the formation of hydrogen bonds with the hydroxyl groups of the $\gamma$-Al$_{2}$O$_{3}$ surface.

The dynamics of glycerol in fully saturated pores is primarly affected by the reduction of mobility at the solid-liquid interface, which is the result of a higher density and structural order in the contact layer than in the bulk liquid and the presence of hydrogen bonds between molecules in the contact layer and hydroxyl groups at the surface. 

Pores that are not fully saturated exhibit a cavity, approximately located in their centre, that does not contain glycerol and causes a disruption of the hydrogen-bond network within the liquid, which in turns enhance the glycerol self-diffusion. The competition between the strong structural order at the solid-liquid interface and the disruption of hydrogen bonds far from it, leads to an overall self-diffusion that is faster than that in the liquid bulk. In particular, we found a significant enhancement in the self-diffusion of nanoconfined glycerol for pore sizes $l_{z} \geq 40$ $\textup{\AA}$ and confinement concentrations $\rho_{\textrm{conf}} \leq 1.0$ g cm$^{-3}$. 

The present study provides a fundamental guideline to understand recent experimental observations on the dynamics of glycerol in confined media \cite{d2012}, which is consistent with the dynamics of glycerol observed in thin films \cite{capponi2010} and highlights the importance of liquid-gas interfaces in the dynamics of confined viscous fluids.

%%%%%%%%%%%%%%%%%%%%%%%%%%%%%%%%%%%%%%%%%%%%%%%%%%%%%%%%%%%%%%%%%%%%%
%% The "Acknowledgement" section can be given in all manuscript
%% classes.  This should be given within the "acknowledgement"
%% environment, which will make the correct section or running title.
%%%%%%%%%%%%%%%%%%%%%%%%%%%%%%%%%%%%%%%%%%%%%%%%%%%%%%%%%%%%%%%%%%%%%

\section*{Conflicts of interest}
There are no conflicts to declare.\\

\section*{Acknowledgements}
The project leading to these results has received funding from the European Union's Horizon 2020 research and innovation programme under the Marie Sk\l{}odowska-Curie grant agreement No 676045 (MULTIMAT). The authors acknowledge the assistance given by IT Services and the use of the Computational Shared Facility at the University of Manchester.\\

%%%%%%%%%%%%%%%%%%%%%%%%%%%%%%%%%%%%%%%%%%%%%%%%%%%%%%%%%%%%%%%%%%%%%
%% The appropriate \bibliography command should be placed here.
%% Notice that the class file automatically sets \bibliographystyle
%% and also names the section correctly.
%%%%%%%%%%%%%%%%%%%%%%%%%%%%%%%%%%%%%%%%%%%%%%%%%%%%%%%%%%%%%%%%%%%%%

\bibliography{campos}

\bibliographystyle{rsc.bst}

%%%END OF MAIN TEXT%%%

%The \balance command can be used to balance the columns on the final page if desired. It should be placed anywhere within the first column of the last page.

\balance

%If notes are included in your references you can change the title from 'References' to 'Notes and references' using the following command:
%\renewcommand\refname{Notes and references}

%%%REFERENCES%%%
%\bibliography{rsc} %You need to replace "rsc" on this line with the name of your .bib file
%\bibliographystyle{rsc} %the RSC's .bst file

\end{document}